\documentclass[10pt,twocolumn,preprintnumbers,amsmath,amssymb,nofootinbib
,superscriptaddress]{revtex4-1}
\usepackage{graphicx,longtable,mathrsfs,color,array}
\usepackage[hidelinks]{hyperref}
\usepackage[usenames,dvipsnames]{xcolor} % colour
\usepackage{amssymb,amsmath,mathtools,mathrsfs,slashed} % maths
\usepackage{epsfig,subfigure,placeins,float} % plots
\usepackage{booktabs,longtable,ctable,multirow} % tables
\usepackage{exscale,relsize} % scale 
\usepackage[normalem]{ulem} % editing
\usepackage{enumerate}
\usepackage{times, mathptmx} % Times font
\usepackage{comment}
\usepackage{color}
\allowdisplaybreaks[1]

  \newcommand{\im}{\rm Im}
  \newcommand{\re}{\rm Re}

\begin{document}
%\title{On the decay of quasi normal modes of a massive scalar field in a Schwarzschild background.} 
\title{Anomalous decay rate of quasinormal modes}
\author{Macarena Lagos}
\email{mlagos@kicp.uchicago.edu}
\affiliation{Kavli Institute for Cosmological Physics, The University of Chicago, Chicago, IL 60637, USA}
\author{Pedro G.~Ferreira}
\email{pedro.ferreira@physics.ox.ac.uk}
\affiliation{Astrophysics, University of Oxford, DWB, Keble Road, Oxford OX1 3RH, UK}
\author{Oliver J.~Tattersall}
\email{oliver.tattersall@physics.ox.ac.uk}
\affiliation{Astrophysics, University of Oxford, DWB, Keble Road, Oxford OX1 3RH, UK}
\date{Received \today; published -- 00, 0000}

\begin{abstract}
The decay timescales of the quasinormal modes  of a massive scalar field  have an intriguing behaviour: they either grow or decay with increasing angular harmonic numbers $\ell$, depending on whether the mass of the scalar field is small or large. We identify the properties of the effective potential of the scalar field that leads to this behaviour and characterize it in detail. If the scalar field is non-minimally coupled, considered here, the scalar quasinormal modes will leak into the gravitational wave signal and will have decaying times that are comparable or smaller than those typical in General Relativity. Hence, these modes could be detectable in the future. Finally, we find that the anomalous behaviour in the decay timescales of quasinormal modes is present in a much larger class of models beyond a simple massive scalar field. 
\end{abstract}

\date{\today}
\maketitle

%%%%%%%%%%%%%%%%%%%%%%%%%%%%%%%%%%%%%%%%%%%%%%%%%%%%%%%%%%%%%%%%%%%%%%%%%%%%%%%%%%%%%%%%%%%%%%%%%%%%%%%%%%%%%%%%%%%%%%%%%%%%%%%%%%%%%%%%%%%%%%%%%%%%%%

\section{Introduction}\label{sec:introduction}
Quasi-normal modes (QNM) are the resonant decaying modes of a black hole (BH) as it settles down to its final stationary state. These modes have an infinite discrete spectrum of complex frequencies, $\omega=\omega_R+i \omega_I$, whose real part $\omega_R$ determines the oscillation timescale of the modes, whereas the complex part $\omega_I$ determines their exponential damping timescale (see e.g.~\cite{Kokkotas:1999bd, Ferrari:2007dd, Berti:2009kk} for reviews on QNM modes).

In General Relativity (GR), QNM are characterized solely by the mass, $M$, and angular momentum, $J$, of the black hole, the angular harmonic indices $(\ell,m)$ and the degree of the harmonic overtone number $n$ \cite{Cardoso:2016ryw}. If there are additional fundamental fields present in the universe, they can affect the QNM spectrum of BHs, and hence QNM can be a powerful tool to test the properties of gravity. Indeed, while a number of tests of gravity only probe the stationary properties of the spacetime (e.g.~orbiting stars around BHs \cite{Will:2007pp,Psaltis:2015uza, Hees:2017aal}, or BH imaging \cite{Psaltis:2010ca, Loeb:2013lfa,deRham:2011by}), QNM carry information on the dynamical regime of gravity and thus provide crucial new information. In particular, in the case of a massive scalar field conformally coupled to gravity, the BH stationary solution is a Kerr metric, as in GR, while the QNM spectrum will generically be given by a linear superposition of the normal GR spectrum and the scalar field spectrum \cite{Tattersall:2017erk}. Studying this scalar field spectrum then will help understand one of the possible deviations from GR that can be tested in the future with a detection of QNM.

The QNM of a scalar field with mass $\mu$ in the presence of a Schwarzschild BH have been studied with numerical and semi-analytical methods \cite{Konoplya:2004wg, Konoplya:2006br,Dolan:2007mj, Tattersall:2018nve}, where it has been found that, at least for the overtone $n = 0$, if we have a light scalar field, then the longest-lived quasinormal modes are those with a high angular number $\ell$, whereas for a heavy scalar field the longest-lived modes are those with a low angular number $\ell$. To elaborate slightly, this means that for small $\mu$, the scalar field fluctuations of the BH become fuzzier (albeit with exponentially fast decaying amplitudes) as it evolves towards its final, perfectly smooth, spherically symmetric state. In contrast, for large $\mu$,  the scalar field fluctuations become smooth, with low $\ell$ modes dominating and eventually disappearing as the BH settles down. 

\begin{figure}
	\centering
	\includegraphics[scale=0.42]{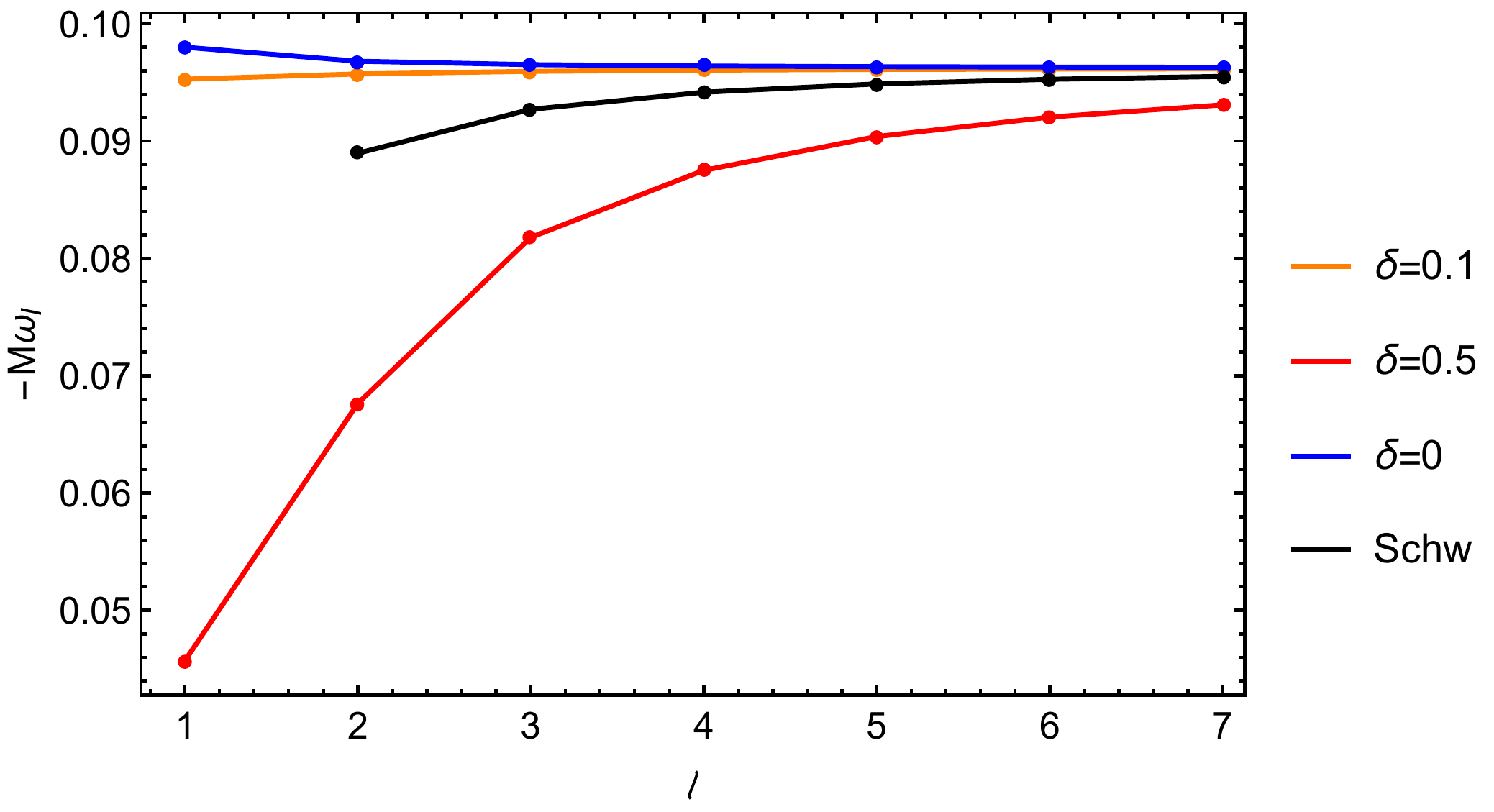}
	\caption{Complex QNM frequency for the cases of a massless scalar field $\delta=0$ (blue), for a massive scalar field with $\delta=$0.1 (orange) and 0.5 (red), and for the tensor modes of Schwarzschild \cite{Berti:2009kk} (black), as a function of the angular number $\ell$. We show the fundamental overtone $n=0$.}
	\label{Fig:Imw_Comparison}
\end{figure}

This peculiar result should be compared to the Schwarzschild tensor QNM spectrum in GR (and the same behaviour can be found for tensor QNM in  Kerr\footnote{From the publicly available data in \cite{QNMdata}, one can check that, for any given azimuthal number $m$, the QNM are such that $|\omega_I|$ grows with $\ell$ for all values of the BH's dimensionless spin $a=J/M^2$.}), where the longest-lived modes are always the ones with lower angular number $\ell$, very much as in the case of a very large $\mu$. Indeed, this is the typical hierarchy found in physical scenarios, where more energetic modes have faster decaying rates (e.g.~electrons in higher energy states in an atom have faster spontaneous decay rates to the ground state, see also examples mentioned in \cite{Maggiore:2007nq}). 
The main goal of this paper will then be to analyze the origin for this ``inversion'' of the decay rate for small $\mu$. 
We compare the behaviour of $\omega_I$ for the massive scalar field and for the tensor modes of Schwarzschild in Fig.~\ref{Fig:Imw_Comparison}. As we will show later, when $\delta=M\mu<0.07$ then $|\omega_I|$ will decay with larger $\ell$. 
On the other hand, we mention that the real part of the scalar QNM frequency, $\omega_R$, does monotonically increase with the angular number $\ell$, as one would expect since higher angular numbers correspond to more energetic systems, and in typical classical and quantum systems the oscillation frequency increases with energy. 

In this paper, we analyze and characterize the non-monotonic behaviour of $\omega_I(\ell)$ for a massive scalar field, focusing mainly on a spherically-symmetric spacetime, but we obtain similar results around a Kerr black hole. We find the critical mass of the scalar field $\hat{\delta}$ where the transition in behaviour happens: for lower masses the large-$\ell$ modes live the longest, whereas for larger masses the small-$\ell$ modes live the longest. 
Using the WKB approximation to analyze the QNM frequencies, we also find that $\omega_I$ is closely related to the width of the effective potential in the QNM equations. We show that the width has a similar non-monotonic behaviour to $\omega_I$ in the case of a massive scalar field. We also find that the presence of this inverted behaviour is common to a broader class of models and not only massive scalar fields.

This paper is organized as follows. In Section \ref{Sec:MassiveScalar} we introduce and discuss the massive scalar field equation. In Section \ref{Sec:WKB} we use the WKB approximation to obtain analytical expressions in the large-$\ell$ limit for the QNM frequencies. We find the critical $\delta$ where $\omega_I$ inverts its behaviour and discuss its observational consequences. In Section \ref{Sec:potentialshape} we study the geometrical properties of the effective potential in the QNM equations and show that the width of this potential causes the anomalous behaviour in $\omega_I$. In Section \ref{Sec:otherpotentials} we show that this anomalous behaviour happens for a range of potentials beyond a simple massive scalar field. 
Finally, in Section \ref{Sec:Summary} we summarize our findings and discuss their consequences. Throughout this paper we will be using geometric units, in which $G=c=\hbar=1$.

%%MASSIVE SCALAR FIELD
\section{Massive Scalar Field}\label{Sec:MassiveScalar}
Scalar fields have been extensively studied in the context of quantum-gravity candidates (e.g.~\cite{METSAEV1987385, Arvanitaki:2009fg}), cosmology (e.g.~as inflatons \cite{Cheung:2007st}, as dark matter \cite{Hu:2000ke}, or dark energy \cite{Gubitosi:2012hu}), and in the strong-field regime as fields modifying the BH spacetime geometry \cite{Herdeiro:2015waa, Silva:2017uqg}, or forming clouds through accretion or instabilities around BHs \cite{Brito:2015oca, Hui:2019aqm, Clough:2019jpm}.

One can go beyond simple scalar field actions and consider non-minimal interactions with the spacetime metric. These can lead to black hole hair (i.e.~a stationary black hole spacetime differing from Kerr and Schwarzschild) \cite{Sotiriou:2014pfa,Herdeiro:2015waa,Babichev:2017guv}.
However, in \cite{Tattersall:2018map} it was argued that, in the presence of a cosmologically-relevant scalar field with second-order equations of motion that does not modify the propagation speed of gravitational waves (in agreement with GW170817 observations \cite{TheLIGOScientific:2017qsa, 2041-8205-848-2-L12}), black holes have no hair and therefore black hole solutions are simply described by Kerr and Schwarzschild. Explicitly, the scalar-tensor theories with second-order equations of motion satisfying these conditions are given by:
\begin{equation*}
S=\int d^4x\, \sqrt{-g}\left[ G_4(\Phi)R+ G_2(\Phi, X)- G_3(\Phi, X)\Box\Phi  \right],
\end{equation*}
where $G_4$ is an arbitrary function of the scalar field $\Phi$, and $G_2$ and $G_3$ arbitrary functions of $\Phi$ and its kinetic term $X=-(1/2)\nabla_\mu \Phi \nabla^\mu \Phi$. Here, $\Box=\nabla^\mu\nabla_\mu$.
In these models, \cite{Tattersall:2018map} showed that the QNM for a massive scalar field $\Phi$ on a black hole spherically-symmetric spacetime will be determined by a Klein-Gordon-like equation:
\begin{equation}
[\Box +\mu(r)^2]\Phi=0,\label{KGeq}
\end{equation} 
where $\mu$ can depend on space and represents an effective mass term, and where the derivatives $\Box$ are taken with respect to  the black hole spacetime. When the spacetime is spherically-symmetric, the line element is given by the Schwarzschild metric:
\begin{equation}\label{Sch}
ds^2=-f(r)dt^2+\frac{1}{f(r)}dr^2+r^2(d\theta^2+\sin\theta^2d\phi^2)
\end{equation} 
with $f(r)=1-r_s/r$ with $r_s=2M$ the Schwarzschild radius. 

In models with non-minimal interactions between the scalar field and the metric of the form $G_4(\Phi)R$, the gravitational waves $h_{\mu\nu}$ of the spacetime metric will be a linear combination of those in GR and the scalar field solutions \cite{Tattersall:2018nve}:
\begin{equation}\label{Schwarzschild}
h_{\mu\nu}=h_{\mu\nu}^{(GR)}-G_{4,\Phi}(r) g_{\mu\nu}\Phi,
\end{equation}
where $G_{4,\Phi}(r)$ is the derivative of the arbitrary function $G_4 $ with respect to the scalar field $\Phi$ evaluated in the background Schwarzschild solution, and $g_{\mu\nu}$ is the Schwarzschild metric. In this scenario, the QNM spectrum that could potentially be observed is given by $h_{\mu\nu}$ and it would have components oscillating with the GR QNM frequencies, in addition to components oscillating with the scalar QNM frequencies. We emphasize that since the background is a black hole with vanishing curvature, $R=0$, then the conformal coupling $G_4$ plays no relevant role in the Schwarzschild equations of motion. However, at the level of linear perturbations, the perturbed curvature will not vanish and therefore $G_4$ becomes relevant, and determines the mixing between scalar and tensor QNM shown in eq.~(\ref{Schwarzschild}).

Since eq.~(\ref{KGeq}) turns out to describe the QNM modes in a range of scalar-tensor theories, we focus on solving this equation. For different models, $\mu(r)$ will take different forms as a function of $r$. For concreteness, we will consider the case where $\mu$ is a constant in most of this paper but, later, we will show our results hold in a wider range of situations.

On the spacetime (\ref{Sch}), the solution to eq.~(\ref{KGeq}) can always be decomposed into spherical harmonics $Y_{\ell m}$ as:
\begin{equation}
 \Phi=  \sum_{\ell, m} \frac{\varphi_{\ell m} (r)}{r} Y^{\ell m}(\theta,\phi) e^{-i\omega_{\ell m}t},
\end{equation}
where $\omega_{\ell m}$ are the QNM frequencies which, in this case, are independent of $m$. 
The radial part of the scalar field satisfies the following equation:
\begin{equation}
\partial_{r_*}^2 \varphi +W(r_*)\varphi=0,\label{varphieq}
\end{equation}
where $r_*=r+r_s\ln (r/r_s-1)$ (hence $dr_*/dr=(1-r_s/r)^{-1}$) is the tortoise coordinate. Here, and onwards, we omit the explicit harmonic number $\ell$ dependence in $\varphi$ and other quantities. The effective potential $W$ is given by:
\begin{equation}\label{Veq}
W=\omega^2-V, \quad V= \left(1-\frac{r_s}{r}\right)\left( \frac{r_s\sigma}{r^3}+\mu^2 +\frac{\ell(\ell +1)}{r^2}\right),
\end{equation}
where $\sigma=1-s^2$ with $s=0$ being the spin of the particle for the case of a scalar field. 
In order to find the QNM frequencies, one must find all the solutions of $\varphi$ that satisfy purely ingoing boundary conditions at the BH horizon, and purely outgoing conditions at spatial infinity.  
This can be done numerically \cite{Konoplya:2004wg, Konoplya:2006br} or semi-analytically \cite{Tattersall:2018nve}. 

\section{QNM with the WKB method} \label{Sec:WKB}
\begin{figure}[h!]
	\centering
	\includegraphics[scale=0.45]{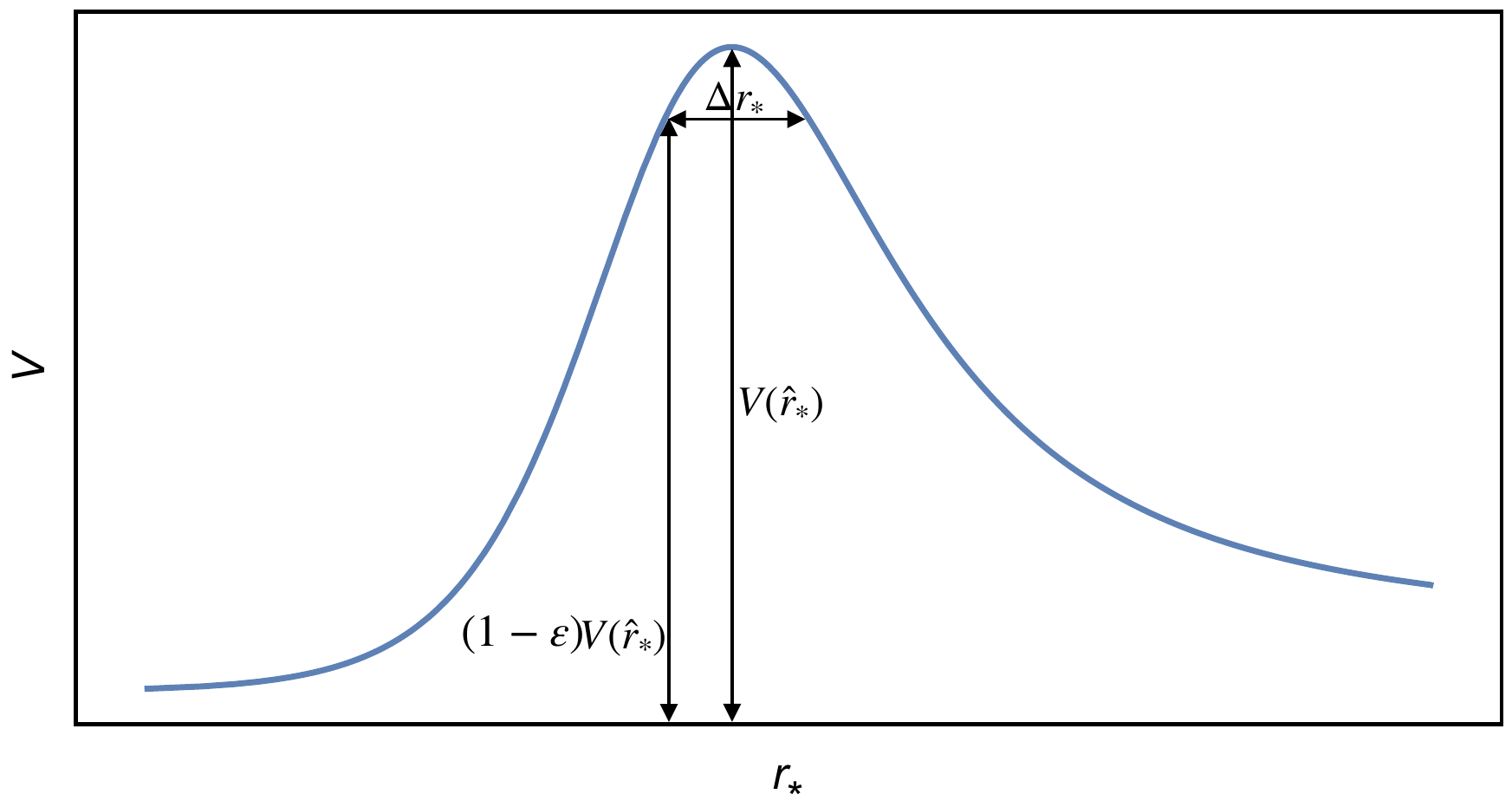}
	\caption{General shape of $V$ as function of tortoise coordinate. We indicate the maximum height $V(\hat{r}_*)$ at $r_*=\hat{r}_*$, as well as the width $\Delta r_*$ of the potential when it has decayed a factor  $(1-\epsilon)$ from its maximum. The height and width play a crucial role in determining the QNM frequencies.}
	\label{Fig:VvsRt}
\end{figure}

The shape of the potential, $V$, as function of the tortoise coordinate is illustrated in Fig.~\ref{Fig:VvsRt}, where we see that the potential tends to constant values at the horizon  $r_*=-\infty$ and spatial infinity $r_*=\infty$, and exhibits a maximum value at a finite distance. 

The overall shape of $V$ is generic for the problems we are considering and makes eq.~(\ref{varphieq}) take the form of a Schr\"{o}dinger equation with a potential barrier $W$.
One can then apply the Wentzel-Kramers-Brillouin (WKB) method as it is used in quantum mechanics to obtain the QNM spectrum. In this case, the QNM frequencies are fully determined by the behaviour of the potential and its derivatives at its maximum value. Given that, it is convenient to  expand the potential around its maximum in a Taylor series as:
\begin{equation}\label{TaylorW}
W(r_*)= \omega^2-V(\hat{r}_*)- \sum_{k=2}^{k=\infty} \frac{1}{k!}V^{(k)}(r_*-\hat{r}_*)^k
\end{equation}
where $V^{(k)}$ denotes the $k$-th derivative of the potential $V$ with respect to the tortoise coordinate, evaluated at the maximum of the potential $\hat{r}_*$. One can truncate this expansion to a given order to obtain a better approximation to the entire potential. The lowest orders have been found to give very accurate results for high-$\ell$ modes. In \cite{1985ApJ...291L..33S,1987PhRvD..35.3621I,Konoplya:2003ii, Hatsuda:2019eoj} the WKB approximation was studied up to 6th order (and higher) in the Taylor expansion (\ref{TaylorW}), where it was found that the frequencies $\omega$ must satisfy a specific constraint in terms of $V$ and the coefficients $V^{(k)}$:
\begin{align}\label{WKBConstraint}
&\omega^2-V(\hat{r}_*)= -2i\left\{   N \sqrt{-V^{(2)}/2}   \right.\nonumber\\
& +\frac{i}{64} \left[  -\frac{1}{9}\frac{V^{(3)2}}{V^{(2)2}}(7+60N^2)+\frac{V^{(4)}}{V^{(2)}}(1+4N^2)\right] \nonumber\\
&  +\frac{N}{2^{3/2}288 }\left[ \frac{5}{24}\frac{V^{(3)4}}{(-V^{(2)})^{9/2}}(77+188N^2) \nonumber\right. \\  
&+ \frac{3}{4}\frac{V^{(3)2}V^{(4)}}{(-V^{(2)})^{7/2}}(51+100N^2)  +\frac{1}{8}\frac{V^{(4)2}}{(-V^{(2)})^{5/2}}(67+68N^2) \nonumber\\
&\left.\left. + \frac{V^{(3)}V^{(5)}}{(-V^{(2)})^{5/2}}(19+28N^2)  
+ \frac{V^{(6)}}{(-V^{(2)})^{3/2}}(5+4N^2)
\right]\right\},
\end{align}
where $N=n+1/2$ with $n=0,1,2,\dots$ the overtone number. 

It is useful to define the function $U$ such that:
\begin{equation}
\omega^2-V(\hat{r}_*)=-2iU,
\end{equation}
where $-2iU$ is given by the RHS of eq.~(\ref{WKBConstraint}), and can therefore be a complex number. The complex part of the QNM frequency, $\omega_I$, can then be expressed as:
\begin{equation}
2 \omega_I^2= -(2\im(U)+V)+\sqrt{(2\im(U)+V)^2+4\re(U)^2} 
\end{equation}
where all the RHS must be evaluated at the maximum $\hat{r}_*$. Meanwhile, the real part of the frequency can be expressed as:
\begin{equation}
\omega_R^2=\re(U)^2/\omega_I^2.
\end{equation}
Note that, by construction, both expressions for $\omega_I^2$ and $\omega_R^2$ are always ensured to be positive.

\subsection{Eikonal Limit}
The eikonal limit of the QNM expansion is given by the limit $\ell\rightarrow \infty$. It has been shown in the past \cite{Cardoso:2008bp,Konoplya:2017wot} that, in some cases, this regime can be associated to the behaviour of null particles trapped on the unstable circular geodesic of spacetime, with $\omega_R$  determined by the angular velocity at the unstable null geodesic, and $\omega_I$ determined by the instability timescale of the orbit. The QNM then represent waves orbiting in the vicinity of the photon ring, slowly leaking out towards the event horizon as well as spatial infinity \cite{1972ApJ...172L..95G}. 

As previously mentioned, the behaviour of the potential at its maximum fully determines the QNM frequencies. In the case of the conformally-coupled massive scalar field (that is, with $G_4\not=0$), this agrees with the null geodesic picture described above, as one can easily verify that the maximum of the potential in eq.~(\ref{Veq}) coincides with the photon ring radius $\hat{r}\approx (3/2)r_s$  for $\ell\rightarrow\infty$. In this context, the QNM frequencies are associated to specific geometrical properties of the potential around the light ring. We illustrate this in the eikonal limit, where it is enough to use the WKB constraint in eq.~(\ref{WKBConstraint}) only up to second-order derivatives of the potential.
In this limit we have
\begin{align}
\omega_I^2&\approx  \frac{1}{8}\frac{-V^{(2)}}{V(\hat{r}_*)}, \label{wIEikonal}\\
\omega^2_R&\approx V(\hat{r}_*)^2.
\end{align}
We can see from here that the QNM frequencies are solely determined by the maximum height of the potential, as well as its width around its maximum (see Fig.~\ref{Fig:VvsRt}). Indeed, using the Taylor expansion to second-order derivatives, we identify the width of the potential when it has decayed from its maximum to $\epsilon V(\hat{r}_*)$ as $(\Delta r_*)^2\approx 2(1-\epsilon)V/(-V^{(2)})$. We can then rewrite eq.~(\ref{wIEikonal}) as:
\begin{equation}
\omega_I^2 \approx  \frac{1}{4}\frac{(1-\epsilon)}{(\Delta r_*)^2}. 
\end{equation}
Note that $\epsilon$ is an arbitrary number that determines where the width of the potential is measured, and it is independent of the QNM numbers $\ell$ and $n$. For the approximation to be valid, $\epsilon$ cannot be too small so that the quadratic Taylor expansion of the potential remains valid. 

The fact that $\omega_I$ scales inversely proportional to the width of the potential can be understood by invoking the aforementioned interpretation of QNM as wave packets orbiting the light ring. 
These waves must be able to tunnel through the potential barrier $W$ in order to leak out. This tunneling rate is precisely what the WKB approach proposed in \cite{1985ApJ...291L..33S,1987PhRvD..35.3621I} analyzes to obtain the QNM frequencies. 

One can construct an interesting, but somewhat artificial, toy analogy: the quantum tunneling across a square barrier but in which the incident wave is forced to zero (reminiscent of no outgoing waves from the event horizon nor incoming from infinity) \cite{1987PhRvD..35.3621I}. Strictly this goes beyond the usual applicability of quantum mechanics in that the eigenvalues are complex. Nevertheless, in that case, only a discrete set of complex, ``energy'' levels are permitted in which the imaginary part is indeed inversely proportional to the width of the barrier.

\subsection{Inverted $\omega_I$ behaviour}

Going beyond the eikonal limit, from the potential $V$ in eq.~(\ref{Veq}) we obtain that the maximum of the potential for large $\ell$ is at:
\begin{equation}
\hat{r}\approx  \frac{3r_s}{2}-\frac{r_s}{6\lambda^2}\left(1 -27 \delta^2 \right)+\mathcal{O}(\lambda^{-4}),
\end{equation}
 where $\lambda^2\equiv \ell(\ell+1)$ and $\delta=\mu M$, and the value of the potential at this position is approximately:
\begin{equation}\label{Vexpand}
V(\hat{r}_*)\approx \frac{4}{27}\frac{\lambda^2}{r_s^2}+ \frac{4}{81}\frac{(2+27\delta^2)}{r_s^2}.
\end{equation}
For the higher derivatives of the potential, we find that (except for $V^{(2)}$) only the leading terms when $\ell\rightarrow \infty$ are relevant in determining the QNM frequencies with eq.~(\ref{WKBConstraint}).  The relevant expansions are then given by:
\begin{align}
V^{(2)} (\hat{r}_*)\approx &  -\frac{32}{729}\frac{\lambda^2}{r_s^4} + \frac{64}{6561r_s^4}(-4+27\delta^2)\\
V^{(3)} (\hat{r}_*)\approx &  \frac{128}{6561}\frac{\lambda^2}{r_s^5} \\
V^{(4)} (\hat{r}_*)\approx & \frac{1024}{19683}\frac{\lambda^2}{r_s^6} \\
V^{(5)} (\hat{r}_*)\approx & -\frac{5120}{59049}\frac{\lambda^2}{r_s^7}\\
V^{(6)} (\hat{r}_*)\approx & -\frac{16384}{177147}\frac{\lambda^2}{r_s^8}\label{V6expand}.
\end{align}
In order to illustrate the overall behaviour of the QNM frequencies, we focus on the fundamental mode $n=0$ from now on (higher overtones will exhibit similar characteristics). For this mode, we find that $U$ can be approximated by:
\begin{align}
U (\hat{r}_*)\approx & \frac{2}{27}\frac{\lambda}{r_s^2}+\frac{1}{52488}\frac{1}{\lambda r_s^2}(-749+11664\delta^2)\\
& + \frac{i}{r_s^2} \left( -\frac{65}{1458} +\frac{1}{6561\lambda^2} (-59+378\delta^2)  \right),
\end{align}
and hence the complex part of the frequency is given by:
\begin{equation}\label{Imw}
r_s^2\omega_I^2\approx  \frac{1}{27}  +\frac{1}{52488\lambda^2}\left(137-29160\delta^2\right) .
\end{equation}
From this analytical expression we find that there is indeed a change in behaviour that happens for $\omega_I(\ell)$: there is a critical mass  $\hat{\delta}=\sqrt{137/29160}\approx 0.06854$, such that for lower masses $|\omega_I|$ decays with $\ell$, whereas for larger masses $|\omega_I|$ grows with $\ell$. Note that the perturbative expansion we have performed is valid only when the first term in (\ref{Imw}) dominates, and therefore only when $\delta^2/\lambda^2 \ll 1$ and $\lambda^2\gg1$. 

\begin{figure}[h!]
	\centering
	\includegraphics[scale=0.42]{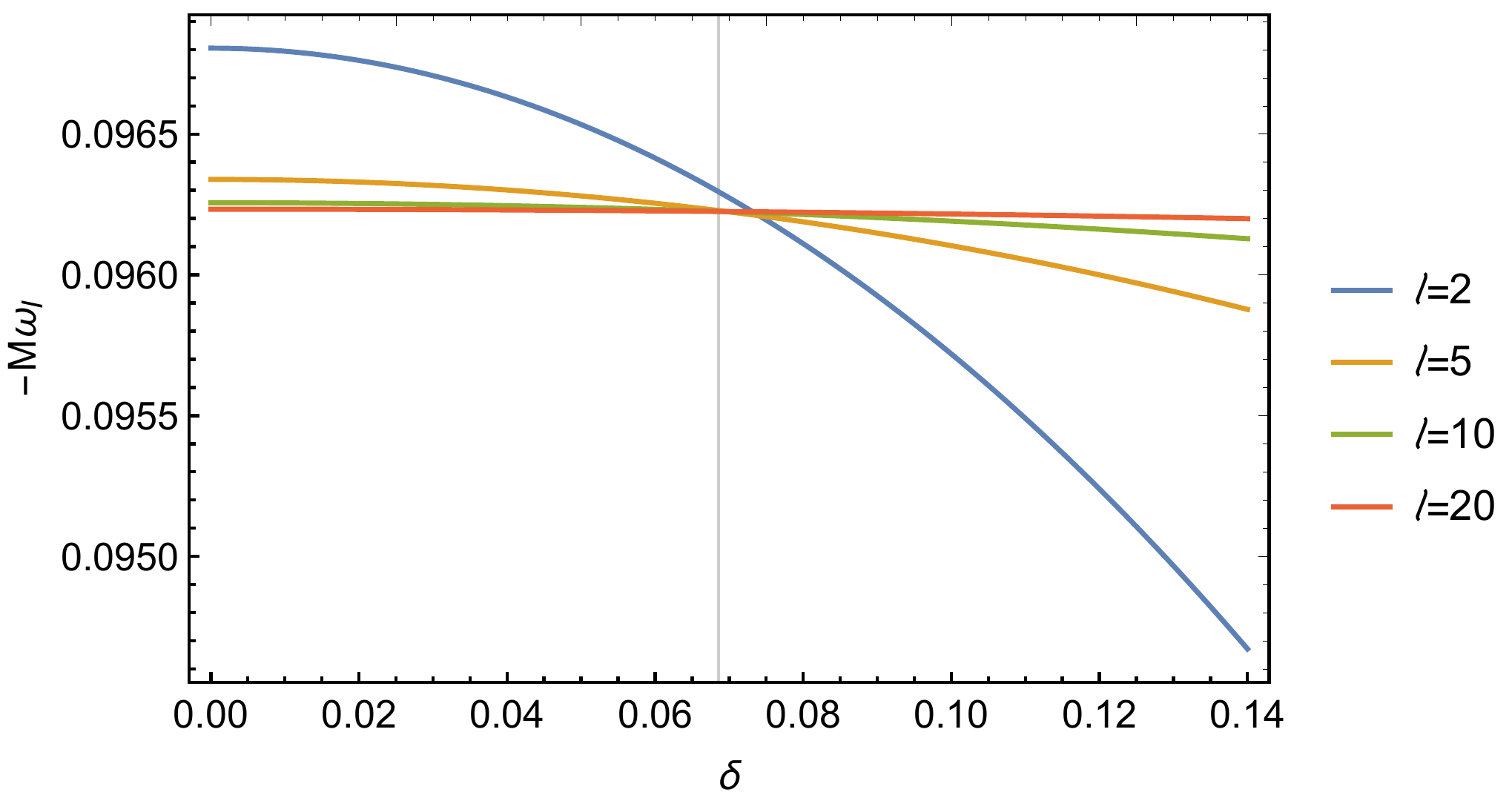}
	\caption{Complex QNM frequency as a function of $\delta$ for a few values of $\ell$. We see that the large-$\ell$ curves cross each other near $\delta\approx 0.068$ (vertical grey line).}
	\label{Fig:Imwvsm}
\end{figure}

For any value of $\ell$, we can also use the WKB expression in eq.~(\ref{WKBConstraint}) and numerically obtain the QNM frequencies. Fig.~\ref{Fig:Imwvsm} shows the numerical results for $\omega_I$ as function of $\delta$ for different values of $\ell$. In this figure we see that all the scalar QNM have $\omega_I<0$ (regardless of their hierarchy in $\ell$), which means that all modes have exponentially fast decaying amplitudes. This ensures the stability of the Schwarzschild black hole under disturbances, as well as the validity of the perturbative approach we have employed. 
We note that for large-$\ell$ values, there is one fixed critical mass $\hat{\delta}$ where curves cross over, but this is not the case for low $\ell$. Indeed, moving beyond the sub-leading term in the eikonal expression for $\omega_I$, we can calculate the critical $\delta$ value for which the imaginary frequency of two QNMs with successive $\ell$ values are equal. That is, using $\omega_I^\ell = \omega_I ^{\ell+1}$ as a proxy for where the transition occurs, we obtain the following expression for $\hat{\delta}$ when using the WKB method:
\begin{align}\label{CriticDelta}
\hat{\delta}^2 &\approx \frac{137}{29160}+\frac{4747969}{566870400}\ell^{-2}-\frac{47479699}{283435200}\ell^{-3}\nonumber\\
&+\frac{295823943463}{11019960576000}\ell^{-4}-\frac{111222908743}{2754990144000}\ell^{-5}.
\end{align}
We note that whilst the results presented so far in this paper have all been obtained using analytical or semi-analytical methods, the calculations are consistent with numerical methods. As an example, we compare to the numerical parameterization of QNMs given in \cite{Cardoso:2019mqo}, whose results were obtained using high accuracy direct integration of QNM equations with generic power law additions to the potential.
For the case of a massive scalar field, \cite{Cardoso:2019mqo} obtained $\hat{\delta}=0.071$ when $\omega_I^{\ell=2}=\omega_I^{\ell=3}$, whereas we obtain $\hat{\delta}=0.072$ using eq.~(\ref{CriticDelta}). There is an even closer agreement when extending eq.~(\ref{CriticDelta}) to further powers in $1/\ell$. 

For a slowly rotating Kerr black hole, we find the following expression for $\hat{\delta}$ using the expressions of \cite{Tattersall:2018nve} for maximally co-rotating modes (those with $m=\ell$) by requiring that $\omega_I^\ell = \omega_I ^{\ell+1}$:
\begin{align}
\hat{\delta}^2 \approx & \; \hat{\delta}^2_0+ \frac{a}{\sqrt{3}}\left[\frac{223}{36450}-\frac{223}{48600}\ell^{-1}+\frac{63240251}{12754584000}\ell^{-2}\right.\nonumber\\
&\left. - \frac{429752621}{51018336000}\ell^{-3} + \frac{6855418935097}{991796451840000}\ell^{-4} \right],
\end{align}
where $a$ is the dimensionless angular momentum of the black hole, and $\hat{\delta}_0$ is the critical value for a Schwarzschild background (i.e.~with $a=0$). We thus find that the critical value $\hat{\delta}$ increases with increasing black hole spin for these modes with $\ell=m$.

Moving beyond the slow rotation approximation, we see this behaviour persist for black holes with significant spin. For example, with $a=0.5$ we find a hierarchy inversion occurs for $\delta \approx 0.09$ when calculating QNMs using Leaver's continued fraction method. The QNM for extremal Kerr BHs have been studied in \cite{Dolan:2007mj, Hod:2011zzd}.

We also mention that the QNM of a massive scalar field around a spherically symmetric wormhole have been studied in \cite{Churilova:2019qph}, where the same anomalous behaviour of $\omega_I(\ell)$ has been found. 

In order to understand better the consequence of the anomalous behaviour in $\omega_I$, let us notice that we can rewrite $\delta=r_s/(2\lambda_c)$, where $\lambda_c$ is the Compton wavelength of the scalar. The transition then happens when $r_s\sim 0.1 \lambda_c$. For a solar mass BH, we will have a transition for $\lambda\sim 10$km (or a scalar mass of $\mu\sim 10^{-10}$eV). We also notice that for $\delta<\hat{\delta}$ we always have that $|\omega_I|$ will be larger than that for Schwarzschild $|\omega_I^{(GR)}|$. Therefore, in all the situations when the scalar QNM have an inverted behaviour in $\ell$, the QNM of the metric will live longer, although the timescales are very similar (see Fig.~\ref{Fig:Imw_Comparison}). In addition, their quality factors, defined as $Q\equiv |\omega_R/(2\omega_I)|$, will also be very similar, with the $Q$ of the scalar always being slightly larger than that of the Schwarzschild QNM. We thus conclude that it could indeed be possible to detect modes with the inverted behaviour, if the sensitivity of gravitational-wave detectors is enough to observe those in GR (assuming that the initial amplitude of both scalar and tensor modes is equally relevant, which must be verified modeling numerically the merger of compact objects in scalar-tensor theories) \cite{Tattersall:2019pvx}.  

We also notice that, for any given $\ell$, $|\omega_I|$ decays with the mass of the scalar field. This means that if the scalar is heavy enough, the scalar QNM modes may live considerably longer than the Schwarzschild ones, and they would dominate the gravitational wave signal. The red curve of Fig.~\ref{Fig:Imw_Comparison} illustrates this for $\delta=0.5$.
For even higher masses, $\omega_I$ will eventually vanish, at which point these modes will disappear from the QNM spectrum \cite{Ohashi:2004wr}.

As also shown in Fig.~\ref{Fig:Imw_Comparison}, we see that in the large-$\ell$ limit, $\omega_I$ converges to the same value as in Schwarzschild, and becomes independent of $\delta$ (since the presence of the mass term $\mu$ in the effective potential $W$ makes a sub-dominant contribution in the limit $\ell\rightarrow \infty$). Indeed, the QNM frequencies for the scalar field will have the same limit as the GR Schwarzschild's QNM:
\begin{equation}
\lim_{\ell\rightarrow\infty} M\omega = \frac{1}{\sqrt{27}}\left[1-i\left(n+\frac{1}{2}\right)\right]
\end{equation}
for any overtone $n$.

Finally, we mention that this anomalous behaviour in $\omega_I$ appears for higher overtones as well, although the critical scalar mass where the inverted behaviour starts will vary with $n$. Explicitly, we find that, for large $\ell$,
\begin{equation}
\hat{\delta}^2= \frac{(137 + 235 n + 235 n^2)}{29160},
\end{equation}
and thus the critical mass grows with $n$.

\section{Potential Shape}\label{Sec:potentialshape}

Let us assume we know exactly what the eikonal limit of the potential  $V$ is, and that only the sub-leading corrections to the potential and its derivatives are unknown.  We can write the coefficients in the Taylor series of the potential in the large-$\ell$ limit as:
\begin{equation}
r_s^{k+2}V^{(k)}\equiv v_{ka}\lambda^2+v_{kb}
\end{equation}
with $V^{(0)}=V(\hat{r})$, and where the coefficients $v_{ka}$ can be read off from eqs.~(\ref{Vexpand})-(\ref{V6expand}) for $n=0$.
We then have:
\begin{align}\label{OmegaIv2b}
r_s^2\omega_I^2= \frac{1}{27} + \frac{1}{\lambda^2}\left(- \frac{295}{52488}-\frac{v_{0b}}{4}-\frac{27v_{2b}}{32} \right).
\end{align}
From here we see that all the sub-dominant terms have a negative sign, which means that if both $v_{0b}$ and $v_{2b}$ are positive then $\omega_I$ will always grow with larger $\ell$, as is typically the case for most examples considered in the literature. However, in the case of a scalar field, $v_{2b}$ can take negative values and dominate the entire sub-leading term, and eventually lead to an inverted behaviour in $\omega_I(\ell)$.  

A necessary condition for the presence of the inverted behaviour in $\omega_I$ is that there exists a $\delta$ (or any other free parameter in the potential $V(r)$, in more general cases) such that:
\begin{equation}\label{DeltaCritic}
-\frac{295}{52488} = \frac{v_{0b}}{4}+\frac{27v_{2b}}{32}.
\end{equation} 
We note that the number in the LHS of eq.~(\ref{DeltaCritic}) is crucial in determining where exactly the inverted behaviour happens. With the WKB approach we have found that this number depends on all the derivatives of the potential (up tho 6th order) at the maximum of the potential in the eikonal limit in a non-trivial way (see Appendix \ref{App:SymbolicOmegaI} for the full symbolic expression), which is the reason why $\hat{\delta}$ has an unnatural value. Indeed, we have verified that if the WKB taylor series is truncated to, for instance, second order in derivatives of the potential, one does not see the inverted behaviour in $\omega_I(\ell)$ as this number is not reproduced correctly. 

\begin{figure}[h!]
	\centering
		\includegraphics[scale=0.47]{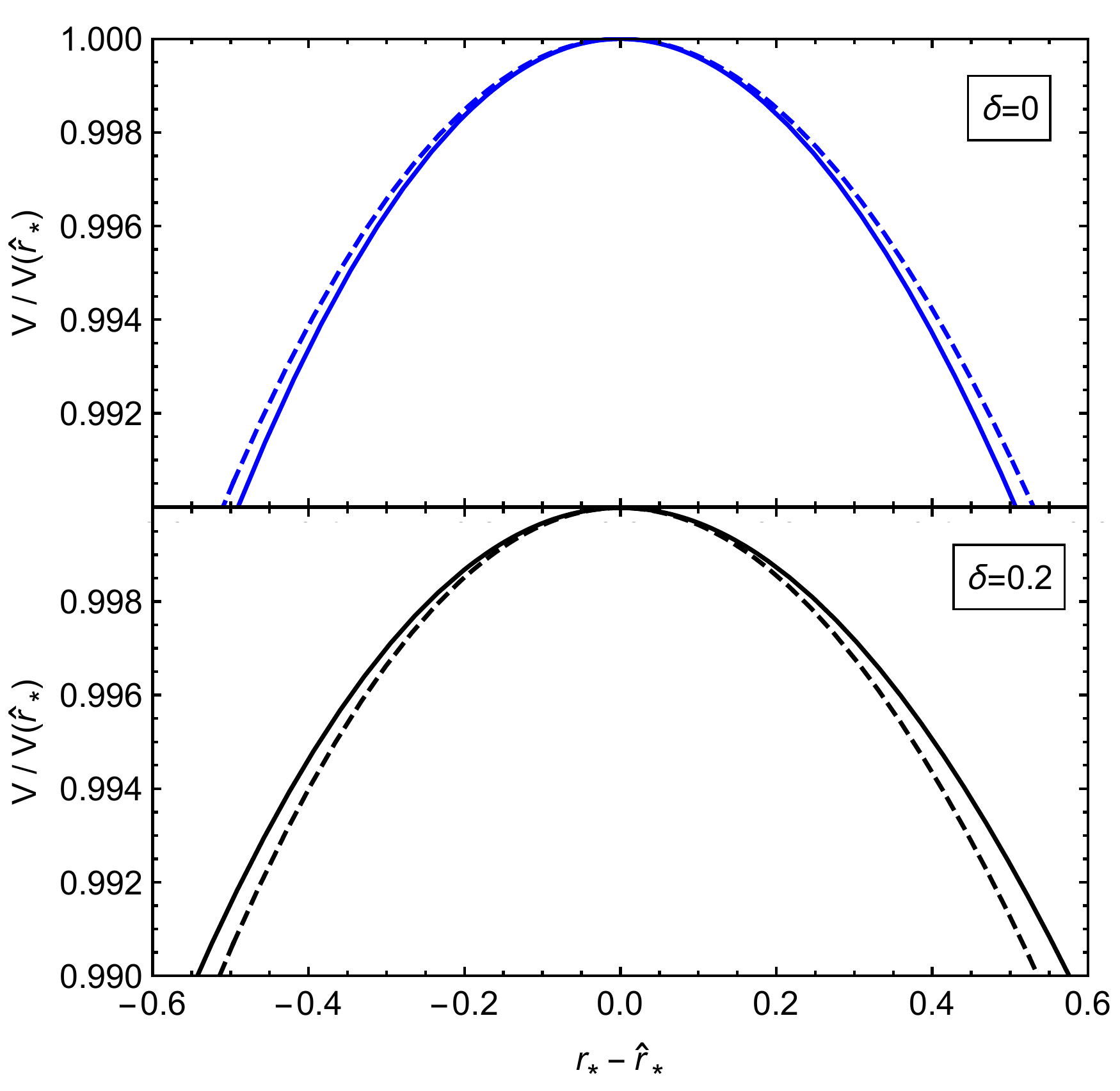}
	\caption{Potential as function of tortoise coordinate for $\delta=0$ (top panel) and $\delta=0.2$ (bottom panel). In both panels the solid line corresponds to $\ell=1$ and the dashed line to $\ell=5$.}
	\label{Fig:width}
\end{figure}

Equation (\ref{OmegaIv2b}) can be re-expressed in the following way:
\begin{equation}
\omega_I^2=-\frac{1}{8}\frac{V^{(2)}}{V(\hat{r}_*)}-\frac{1}{r_s^2\lambda^2}\frac{295}{52488}.
\end{equation}
where $V$ and $V^{(2)}$ include their leading and sub-leading order terms when $\ell\rightarrow\infty$. This expression is a generalization of eq.~(\ref{wIEikonal}) that now is valid beyond the leading order of the eikonal limit.
Thus, we find the following relationship between the width of the potential and the complex frequency to sub-leading order:
\begin{equation}\label{OmegaIWidth}
\omega_I^2 \approx  \frac{1}{4}\frac{(1-\epsilon)}{(\Delta r_*)^2} -\frac{1}{r_s^2\lambda^2}\frac{295}{52488}.
\end{equation}
This expression illustrates explicitly the fact that if the eikonal limit of the potential is known, then the only unknown quantity determining the behaviour of $\omega_I$ is the width $(\Delta r_*)^2$. The width is given by:
\begin{equation}
(\Delta r_*)^2\approx -2(1-\epsilon)r_s^2\frac{v_{0a}}{v_{2a}}\left[1+ \frac{v_{0b}}{v_{0a}\lambda^2}-\frac{v_{2b}}{v_{2a}\lambda^2}\right],
\end{equation}
which contains a leading term fixed by the potential in the eikonal limit, as well as sub-leading terms that depend on the behaviour of the potential beyond the eikonal limit. For the case of the scalar field, the width will behave as:
\begin{equation}
(\Delta r_*)^2\approx \frac{3}{4}(1-\epsilon)r_s^2\left[9+\frac{(-2+135\delta^2)}{\lambda^2}\right],
\end{equation}
from which we again see an inverted behaviour. 
We plot the potential for a couple of values of mass and $\ell$ in Fig.~\ref{Fig:width}, and compare the width when $\epsilon=0.01$.
Here we see the shape of the potential in four cases. The potential is nearly symmetric as we are close to its maximum and the quadratic approximation holds. In the top panel we have the potential for $\delta=0$ and in the bottom panel for $\delta=0.2$. In both panels, the solid line corresponds to $\ell=1$ and the dashed line to $\ell=5$. We see that for $\delta=0$ the width grows with $\ell$, whereas for $\delta=0.2$ the width gets smaller with $\ell$.
We note that this inversion in the behaviour of the width does not happen exactly at the same critical value $\hat{\delta}$ where $\omega_I$ exhibits an inverted behaviour, due to the presence of the second term in the RHS of eq.~(\ref{OmegaIWidth}). Nevertheless, the inverted behaviour of $\omega_I$ is caused by the one in $\Delta r_*$.

%Other potentials
\section{Other Potentials}\label{Sec:otherpotentials}
Let us now generalize the potential $V$ to \begin{equation}
V(r)=\left(1-\frac{r_s}{r}\right)\left(\frac{\ell(\ell+1)}{r^2} + V_\text{eff}(r)\right)
\end{equation}
where $V_\text{eff}(r)$ can take different forms depending on the specific model we have in mind, although we assume that it is independent of $\ell$ (so that the eikonal limit of the potential is known).  In this scenario, $V_\text{eff}$ crucially determines the behaviour of the potential in the sub-leading regime of the eikonal limit.
In terms of the potential $V_\text{eff}$, the criticality condition (\ref{DeltaCritic}) can be rewritten as:
\begin{equation}\label{CriticCond}
-\frac{1180}{729}=    40r_s^2\hat{V}_\text{eff} +48 r_s^3\hat{V}_\text{eff}^{'}+9r_s^4\hat{V}_\text{eff}^{''},
\end{equation}
where these derivatives are with respect to $r$ and evaluated at the maximum of the potential $\hat{r}=3r_s/2$. We note that the same critical condition given by eq.~(\ref{CriticCond}) can also be derived by calculating the QNMs using the analytic series expansion method of \cite{Dolan:2009nk}  and looking at the sub-leading term for $\omega_I$ in the eikonal limit. Analytic expressions for the QNMs of massive scalar fields obtained through the series expansion method of \cite{Dolan:2009nk} can be found in \cite{Tattersall:2018nve}.

For a massive scalar field we have that
\begin{equation}
V_\text{eff}(r,\mu)= \frac{r_s}{r^3}+\mu^2
\end{equation}
and therefore
\begin{equation}
r_s^2\hat{V}_\text{eff}= \frac{8}{27}+4\delta^2,\;  r_s^3\hat{V}_\text{eff}'= -\frac{16}{27},\; r_s^4\hat{V}_\text{eff}^{''}=\frac{128}{81}
\end{equation}
and the condition in eq.~(\ref{CriticCond}) lets us recover
\begin{equation}
  29160\delta^2-137 =0
\end{equation}
as before. 

We can use the generic nature of the condition given by eq.~(\ref{CriticCond}) to probe how general a hierarchy inversion of imaginary frequencies can be. If we again use a scalar field as our starting point, we can imagine introducing a simple power law addition to the effective potential:
\begin{align}\label{VScalGen}
V_{\text{eff}}(r)=\frac{r_s}{r^3}+\frac{\alpha_q}{r_s^2}\left(\frac{r_s}{r}\right)^q.
\end{align}
Now, we find that eq.~(\ref{CriticCond}) can be simplified into a condition for a critical value $\hat{\alpha}_q$:
\begin{align}\label{CriticAlpha}
\hat{\alpha}_q=\frac{3^{q-6}137}{2^q(q-5)(q-2).}
\end{align}
We see that there is no defined critical value for $q=2,5$. The $q=2$ case is unsurprising as this simply corresponds to a change in the effective $\ell$ value of each QNM. The special status of $q=5$, however, is currently unclear.

Examples of potentials with the form of eq.~(\ref{VScalGen}) include that of Einstein-dilaton-Gauss-Bonnet gravity, where perturbations to the dilaton field are sourced by the Kretschmann Scalar $K$ of the background spacetime (for Schwarzschild, $K=R_{\mu\nu\alpha\beta}R^{\mu\nu\alpha\beta} = 48M^2r^{-6}$) \cite{Kanti:1995vq,Silva:2017uqg,Konoplya:2019hml}.
 
We emphasize that this transition in the QNM is present also in more complicated models. In \cite{Konoplya:2005hr, Rosa:2011my}, the QNM of a massive vector field in a Schwarzschild black hole was analyzed (and for Kerr in \cite{Pani:2012bp}). A massive vector field generically propagates three degrees of freedom, corresponding to the helicity-0 and $\pm 1$ modes. The QNM equations can be reduced to one equation of the form (\ref{varphieq}) for an odd-parity field, and two coupled equations of similar form for two even-parity modes. The odd-parity mode has the potential in eq.~(\ref{Veq}) with $\sigma=0$ (and hence $V_\text{eff}=\mu^2$). In this case, there is no mass $\mu$ for which the critical condition (\ref{CriticCond}) is satisfied and therefore this mode does not present the inverted behaviour in $\omega_I(\ell)$, which agrees with the numerical results found in Fig.~1 of \cite{Rosa:2011my}. 
Regarding the two even-parity modes, one of them corresponds to a `scalar' wave that has a non-vanishing monopole $\ell=0$ component. This mode is found to have an inverted behaviour in \cite{Rosa:2011my}. Meanwhile, the `vector' even-parity mode does not exhibit such an inverted behaviour.

The QNM of a massive spin-2 particle in a Schwarzschild black hole has been analyzed in \cite{Brito:2013wya}. In this case, there is a total of five degrees of freedom, corresponding to all the possible helicities. The odd-parity sector will have two coupled degrees of freedom (associated to a vector and tensor mode), and the even-parity sector will have three coupled degrees of freedom (a scalar, vector, and tensor mode). Due to the difficulty of the equations,  \cite{Brito:2013wya} analyzed only the QNM for the odd sector where no inverted behaviour was found. Whether this behaviour emerges in the even sector where the `scalar' mode appears is beyond the scope of this paper and remains to be seen.

\section{Conclusions} \label{Sec:Summary}
In this paper we have explored the QNM of a massive scalar field. 
Both in the case of a Schwarzschild and Kerr black hole, we have found that the complex part of the QNM frequencies $\omega_I$, which determines the timescale in which these modes decay, has an intriguing behaviour as a function of the spherical harmonic number $\ell$. In particular, we found that there exists a critical mass of the scalar field such that for lower masses $|\omega_I|$ decays with larger $\ell$, contrary to most cases studied in the literature. On the other hand, for larger masses, $|\omega_I|$ grows with larger $\ell$ and the standard behaviour is recovered. 

Using the WKB method in the case of Schwarzschild, we found that $\omega_I$ is closely related to the width of the effective potential in the QNM equations, which exhibits the same anomalous behaviour as $\omega_I$. Indeed, QNM can be interpreted as wave packets orbiting in the vicinity of the photon ring of the black hole, slowly leaking out towards the event horizon as well as spatial infinity. In order for this to happen, these waves must be able to tunnel through the effective potential that acts as a barrier. A simple analogous toy model of quantum tunneling shows that the energy of the system indeed takes discrete, albeit complex, ``energy'' values in which the imaginary part is inversely proportional to the width of the barrier, as we have found for the QNM of the massive scalar field. 

While we have identified the feature of the potential -- its width -- that dictates how $\omega_I$ behaves and shown how it also undergoes a transition at a particular value of $\delta$, the underlying reason for this transition remains to be understood. Given the simplicity of the set up, the best way forward will be to construct analogous examples that may shed more light on the phenomenon. 

Interestingly, we have found evidence that this transition to the anomalous behaviour occurs in other physical situations. An open question is whether this phenomenon is generic for the zero helicity components of any massive field. Often these modes arise coupled to other degrees of freedom and are difficult to disentangle ({\it vide} the case of spin-2 fields studied in \cite{Brito:2013wya}) and thus more work needs to be done in streamlining such calculations. 

Finally, we emphasize that if a massive scalar field has a non-minimal coupling to the metric, then the scalar QNM will leak to the QNM of the spacetime metric and become potentially observable. Indeed, we showed that if a gravitational wave detector has the sensitivity to detect the standard QNM of a Schwarzschild black hole, then it should also detect the scalar QNM provided they are present and that the initial amplitude of both scalar and tensor modes are equally relevant, which must be verified modeling numerically the merger of compact objects in scalar-tensor theories in the future.

\textit{Acknowledgements ---} We are grateful for discussions with E.~Berti, V.~Cardoso and N.~Yunes. We thank R.~Brito, V.~Cardoso, S.~Dolan, P.~Pani and J.~Rosa for generously supplying us with the data from \cite{Brito:2013wya} and \cite{Rosa:2011my}.  PGF and OJT acknowledge support from the  Beecroft Trust. This  project  has  received  funding  from  the European  Research  Council  (ERC)  under  the  European  Union's Horizon  2020  research  and  innovation  programme  (grant  agreement No 693024)  and from the Swiss National Science Foundation. ML was supported by the Kavli Institute for Cosmological Physics at the University of Chicago through an endowment from the Kavli Foundation and its founder Fred Kavli.

\appendix

\section{Sub-leading order of $\omega_I$}\label{App:SymbolicOmegaI}
If we generically write the coefficients in the Taylor series of the potential in the large-$\ell$ limit as:
\begin{equation}
r_s^{k+2}V^{(k)}=v_{ka}\lambda^2+v_{kb}
\end{equation}
with $V^{(0)}=V(\hat{r})$, then the complex QNM is given at sub-leading order in $\ell$ by:
\begin{align}
&r_s^2\omega_I^2\approx -\frac{1}{8}\frac{v_{2a}}{v_{0a}}
+\frac{1}{8v_{0a}^2 \lambda^2 }\left(   v_{0b} v_{2a}  -  v_{0a} v_{2b} - \frac{1}{8} \frac{v_{2a}^2}{v_{0a}} - \frac{11}{144}  \frac{v_{3a}^2}{v_{2a}} \right. \nonumber\\
&\left. + \frac{155}{1728}\frac{v_{0a}v_{3a}^4}{v_{2a}^4}+
\frac{1}{16} v_{4a}-\frac{19}{96}\frac{v_{0a}v_{3a}^2}{v_{2a}^3}v_{4a}\right.\nonumber\\
&\left. +\frac{7}{192}\frac{v_{0a}}{v_{2a}^2}v_{4a}^2 +\frac{13}{144}\frac{v_{0a}}{v_{2a}^2}v_{3a}v_{5a}-\frac{1}{48}\frac{v_{0a}}{v_{2a}}v_{6a}\right). 
\end{align}
Here we see that the coefficients $v_{ka}$ for all $k$ contribute to the expression (i.e.~all the derivatives of the potential in the eikonal limit), whereas only $v_{kb}$ for $k=(0,2)$ contribute.

\bibliography{RefModifiedGravity} 

 \end{document}